\documentstyle[11pt,epsfig,color]{article}
\begin{document}

\begin{center}
{\Large\bf In search of the dark matter dark energy interaction: a kinematic approach}
\\[15mm]
Ankan Mukherjee, \footnote{E-mail: ankan\_ju@iiserkol.ac.in}~~
Narayan Banerjee \footnote{E-mail: narayan@iiserkol.ac.in}

{\em $^{1,2}$Department of Physical Sciences,~~\\Indian Institute of Science Education and Research Kolkata,~~\\Mohanpur, West Bengal-741246, India}\\[15mm]
\end{center}

\newcommand{\be}{\begin{equation}}
\newcommand{\ee}{\end{equation}}
\newcommand{\ba}{\begin{eqnarray}}
\newcommand{\ea}{\end{eqnarray}}

\begin{abstract}
 The present work deals with a kinematic approach to the modelling the late time dynamics of the universe. This approach is based upon the assumption of constant value of cosmological jerk parameter, which is the dimensionless representation of the 3rd order time derivative of the scale factor. For the $\Lambda$CDM model, the value of jerk parameter is $-1$ throughout the evolution history. Now any model dependent estimation of the value of the jerk parameter would indicate the deviation of the model from the cosmological constant. In the present work, it has also been shown that for a constant jerk parameter model, any deviation of its value from $-1$ would not allow the dark matter to have an independent conservation, thus indicating towards an interaction between dark matter and dark energy. Statistical analysis with different observational data sets (namely the observational Hubble parameter data (OHD), the type Ia supernova data (SNe), and the baryon acoustic oscillation data (BAO)) lead to a well constrained values of the jerk parameter and the model remains at a very close proximity of the $\Lambda$CDM. The possibility of interaction is found to be more likely at high redshift rather than at present epoch.
\end{abstract}

\vskip 1.0cm

PACS numbers: 98.80.Cq;  98.70.Vc

Keywords: cosmology, dark energy, reconstruction, deceleration parameter, jerk parameter, interacting dark energy.

\section{Introduction}
The recent cosmic acceleration has emerged as an intriguing phenomenon in cosmology. It was first observed during supernova observations in late nineties \cite{RiessPerlmutter}. Subsequent cosmological  observations like type Ia supernova data \cite{suzuki,betoul}, galaxy cluster measurement \cite{allensw} etc have confirmed the accelerated expansion of the universe. The discovery of this phenomenon has propelled the research in cosmology towards a new direction, particularly the basic understanding about the contents of the universe has been  changed radically.  The spectacular development in cosmological observations in the last two decades have imposed tighter constraints on various cosmological models, but the genesis of the cosmic acceleration is still far from being resolved.

\par Various possibilities are explored in order to find an explanation of the late time cosmic acceleration. One way is to introduce an exotic component in the matter sector. This exotic component, dubbed as {\it dark energy}, with a  characteristic negative pressure, leads to the repulsive nature of gravity at cosmological scale. The most popular model of dark energy is the cosmological constant $\Lambda$, which is based upon the assumption that the constant vacuum energy density serves as the candidate of dark energy. The appearance of $\Lambda$ in cosmological models is not new and thus a less radical change in the cosmological models. Though strongly supported by observations, the $\Lambda$CDM (cosmological constant $\Lambda$ with pressureless cold dark mattre) model suffers from various inconsistencies, mainly the fine tuning problem. An elaborate discussion on the merits and problems of the  $\Lambda$CDM model and various related issues have been discussed by Padmanabhan\cite{padmanabhan}. Dark energy candidates that evolve with time  provide efficient alternatives, but have the disadvantage of not properly motivated from other branches of physics.  There  are some excellent reviews on different dark energy models\cite{sahnicopeland}. The other way to look for a plausible reason behind the cosmic acceleration is to find a suitable modification of the General Relativity (GR) such as $f(R)$ gravity models \cite{CappozilloCarroll}, scalar-tensor theories \cite{scalten} and different higher dimensional gravity theories \cite{dvalideffayet}. The most common problem of such models is that they can hardly match GR in the context of local astronomy. 

\par The recent trend in the modelling of cosmic evolution is to build up the model from observational data. This reverse way of finding viable a cosmological model is called {\it reconstruction}. Pioneering work in this direction was by Starobinsky \cite{starore}, where the scalar field potential, used as the dark energy, has been reconstructed by using the density perturbation data. The data of distance measurement of supernova has been utilized in the context of reconstruction by Huterer and Turner \cite{huterner} and by Saini et al. \cite{sainiray}. Parametrization of quintessence scalar field and potential from effective equation of state of dark energy has been discussed by Guo, Ohat and Zhang \cite{guohta}. Construction of non-canonical kinematic terms has been discussed by Li, Guo and Zhang \cite{liguo}. There are two types of practice in reconstruction. The first one is parametric reconstruction which is based upon the assumption of a parametric form of cosmological quantities  like the dark energy equation of state ($w_{DE}$), dark energy density ($\rho_{DE}$), the quintessence the potential etc \cite{cpl,gph} and an estimation of the parameters from the available data. The other one is a non-parametric reconstruction which is an attempt to estimate the evolution of $w_{DE}$ directly from observational data without an assumption of any parametric form \cite{nonpara}.

\par The normal practice in cosmology is to write down Einstein equations, $G_{\mu\nu}=-8\pi G T_{\mu\nu}$ for a spatially homogeneous and isotropic model with the right hand side taking care of the matter sector. The present work has a completely different approach. We assume a spatially flat homogeneous and isotropic metric and define the usual kinematical quantites like the Hubble paramater $H$, the deceleration parameter $q$ and the jerk parameter $j$, which are respectively the first order, second order and third order time derivatives of the scale factor $a$. The derivatives are all fractional derivatives and furthermore $q$ and $j$ are dimensionless. We have observational results of the evolution of $q$, in the sense that the parameter is negative at the present epoch and had been positive in a recent past, the epoch of transition from the decelerated to the accelerated expansion is also known. The natural choice of the kinematical quantity of interest is thus the evolution of $q$, which is the jerk parameter $j$. We now assume a constant jerk and find the evolution of the other kinematical quantities from the definition of jerk. The values of the various kinematical quantities and the model parameters, which come out as the constant of integration and the value of $j$, are then estimated from known observational data sets. 

\par The reconstruction technique normally involves finding out the equation state parameter $w_{DE}$ given by $w_{DE}=\frac{p_{DE}}{\rho_{DE}}$, the ratio of the contribution to the pressure and density sectors from the dark energy. The approach is indeed physical, as it directly talks about the nature of the dark energy sector. Although much less used, the kinematical approach has the virtue of not having any apriori prejudice regarding the dark energy. Pioneering work in the kinematic modelling of cosmic acceleration was by Riess {\it et al.} \cite{riess} where a linear parametrization of the deceleration parameter has been adopted to estimate the redshift at which the transition from decelerated to accelerated expansion phase occurred. Different parametrization of deceleration parameter have been discussed by Shapiro and Turner \cite{shapiroturner}, Gong and Wang \cite{gongwang}, Xu and Liu \cite{xuliu} and Elgaroy and Multamaki \cite{elgaroymultamaki}. Recently kinematic method to investigate the cosmic acceleration has been discussed by Barboza and Carvalho \cite{barbozacarvalho}. 

\par As already mentioned, the jerk parameter had hardly been used until very recently. However, its importance in building a cosmological model had been emphasized long back in terms of a ``state-finder'' parameter\cite{varun1}. The indication of importance of jerk as a future tool for the reconstruction of cosmological models was also indicated by Alam, Sahni, Saini and Starobinsky\cite{varun2}. The reason for its being overlooked as the starting point of reconstruction was perhaps the unavailability of clean data. Reconstruction of dark energy model using deceleration parameter ($q$) and jerk ($j$) as model parameters was discussed by Rapetti {\it et al.} \cite{rapetti} where the present values of the kinematical parameters have been constrained using observational data. Parametrization of time evolving jerk parameter models  have been discussed recently by Zhai {\it et al.} \cite{zhai} and by Mukherjee and Banerjee \cite{mukban}. 

\par If the agent driving the present acceleration is the cosmological constant, then certainly the dark matter sector follows its own conservation equation. However, if the dark energy is an evolving one, there is always a possibility that the two dark sectors interact with each other, one may grow at the expense of the other. Naturally there is a lot of work in the literature where the interacting dark energy model has been discussed. Cosmological evolution of interacting phantom dark energy has been discussed by Guo, Cai and Zhang \cite{guocai}. Guo, Ohta and Tsujikawa have emphasised on the observational constraints on the coupling between different dark components of the universe \cite{guohtatsu}. The last two investigations assume that the interaction is proportional to the total energy density. An interacting dark energy model has recently been  given by Pan, Bhattacharaya and Chakraborty \cite{panbhatta} where again the interaction term is assumed to be proportional to the total energy density. Holographic dark energy models with Hubble scale as the infra red cut off require the interaction between dark energy and dark matter to yield the recent accelerated expansion with a history of a decelerated expansion in the past. Interacting holographic dark energy model has been discussed by Zimdahl and Pavon \cite{zimdahlpavon}. For a {\it graceful entry} of the universe from a decelerated to an accelerated phase in Brans-Dicke theory, the interaction of the Brans-Dicke scalar field and the quintessence scalar field had been discussed by Das and Banerjee \cite{sudipta}. An attempt towards a covariant Lagrangian formulation of the interaction has been made by Faraoni, Dent and Saridakis\cite{faraoni}. Reconstruction of the interaction rate in holographic dark energy model has been discussed by Sen and Pavon \cite{senpavon} where the reconstruction has been done with a prior assumption about the dark energy equation of state. Reconstruction of interaction rate in holographic dark energy from parametrizations of deceleration parameter has been discussed by Mukherjee \cite{amholography}. Recently non-parametric reconstruction of dark energy interaction using Gaussian process has been discussed by Yang, Guo and Cai \cite{yangguocai} where the signature of dark energy interaction has been obtained for a deviation of dark energy equation of state parameter from the value -1.  

\par The present work is an attempt to reconstruct the possible interaction of various matter components from the data sets in a kinematical approach. The starting point is a constant jerk parameter. The result is that any deviation from the $\Lambda$CDM model indicates a possibility of an interaction amongst various matter sectors. The best fit values, however, are tantalizingly close to the $\Lambda$CDM scenario. Another important result is that the allowance of any interaction is more stringent at recent times, but slightly more relaxed in the past.

\par It should also be mentioned at the outset that the entire work depends upon the dogma that a $\Lambda$CDM model should be included as a possibility in an endeavour leading to the reconstruction of the dark energy, at least as a limit.

\par In the following section (section II), the reconstruction of the model has been discussed. The results of the statistical analysis have been presented in section III. A discussion of the results and some concluding remarks have been included in section IV.

\section{Reconstruction of the model for a constant jerk parameter}
The mathematical framework of cosmology begins with the assumption of a homogeneous and isotropic universe, where the distance element is defined as 
\be
ds^2=-dt^2+a^2(t)\Big[\frac{dr^2}{1-kr^2}+r^2d\Omega^2\Big].
\ee 
Here $k$, which can have values $0$ or $\pm1$, conveys the information regarding the nature of spatial curvature. The time dependent quantity in the coefficient of the spatial part of the metric, $a(t)$, is called the scale factor. It takes care of the time evolution of the spatial separation of two space-time point. Here further calculations have been continued assuming the spatial flatness of the universe, i.e. $k=0$.    
\par The fractional rate of expansion of the linear size of the universe, dubbed as the Hubble parameter, is defined as
\be
H(t)=\frac{\dot{a}}{a},
\label{hubbleparameter}
\ee
where the overhead dot denotes a derivative with respect to the cosmic time $t$. To understand the nature of the expansion, higher order time derivatives of the scale factor are to be invoked. The measure of cosmic acceleration is presented in a dimensionless way by the deceleration parameter $q$, defined as
\be
q(t)=-\frac{\ddot{a}/a}{\dot{a}^2/a^2}=-1-\frac{\dot{H}}{H^2}.
\label{decelerationparameter}
\ee
If the value of the deceleration parameter is negative, then the expansion is accelerated.
\par The cosmic `jerk parameter', which is the dimensionless representation of the third order time derivative of the scale factor, is defined as
\be
j(t)=-\frac{1}{aH^3}\Bigg(\frac{d^3a}{dt^3}\Bigg).
\label{jerkparameter}
\ee
  
\par It is convenient to convert the time derivatives to the derivatives with respect to the redshift $z$ (where $1+z=a_0/a$, $a_0$ being the present value of $a$) for studying the dynamics of the universe as $z$ is a dimensionless quantity. From the equation (\ref{jerkparameter}), the expression for the jerk parameter will be
\be
j(z)=-1+(1+z)\frac{(h^2)'}{h^2}-\frac{1}{2}(1+z)^2\frac{(h^2)''}{h^2},
\label{jerkequation}
\ee   
where $h(z)=\frac{H(z)}{H_0}$, ($H_0$ being the present value of the Hubble parameter) and a prime denotes the derivative with respect to $z$. In the present work, the reconstruction is done with the assumption that $j$ is a slowly varying quantity, and will be considered a constant in the subsequent discussion. The solution of the differential equation (\ref{jerkequation}) yields the expression of $h^2(z)$ as
\be
h^2(z)=A(1+z)^{\frac{3+\sqrt{9-8(1+j)}}{2}}+B(1+z)^{\frac{3-\sqrt{9-8(1+j)}}{2}},
\label{h2zABj}
\ee
where $A$ and $B$ are the constant dimensionless coefficients. Now the relation between $A$ and $B$ is obtained from the boundary condition $h(z=0)=1$ as $A+B=1$. Finally $h^2(z)$ is written as a function of redshift $z$ and two parameters $j$ and $A$ as
\be
h^2(z)=A(1+z)^{\frac{3+\sqrt{9-8(1+j)}}{2}}+(1-A)(1+z)^{\frac{3-\sqrt{9-8(1+j)}}{2}}.
\label{h2zAj}
\ee
Therefore this is effectively a two parameter model where $j$ and $A$ are the model parameters. The value of $j$ obtained from the statistical analysis of the reconstructed model using different observational data would indicate the consistency or deviation of this model from the $\Lambda$CDM and it exactly mimics the $\Lambda$CDM for $j=-1$. The deceleration parameter (defined in equation (\ref{decelerationparameter})) can also be expressed for the present model in terms of the model parameters and the redshift as,

\be
q(z)=-1+\frac{A\Bigg(\frac{3+\sqrt{9-8(1+j)}}{4}\Bigg)(1+z)^{\frac{3+\sqrt{9-8(1+j)}}{2}}}{h^2(z)}
+\frac{(1-A)\Bigg(\frac{3-\sqrt{9-8(1+j)}}{4}\Bigg)(1+z)^{\frac{3-\sqrt{9-8(1+j)}}{2}}}{h^2(z)}.
\ee

\par One component of the matter content of the universe, whether it interacts with the dark energy sector or not, is generally believed to be a cold dark matter with an equation of state $p=0$. If we stick to this presupposition, and attempt to recover a non-interacting pressureless fluid at least as a limit from equation (\ref{h2zAj}) for some value of $j$, we find that the second term of the right hand side of equation (\ref{h2zAj}) can yield a highest power of $(1+z)$ as $3/2$ and can not serve the purpose. The only possibility that remains is $j=-1$ which yield the standard $(1+z)^3$ behaviour in the first term. So we identify the first term to represent the contribution from the cold dark matter, which, in the non-interacting limit, yield a $(1+z)^3$ behaviour as in the standard dust model. The rest of the work will depend on this identification. One should note that this is not the only plausible choice. It may well be possible to find a corresponding pressure to each of the contribution to the matter sector so that both the component conserve by themselves. One can easily calculate the equation of state parameter $w$ (given by $w=\frac{pressure}{density}$) for both the contribution. A straightforward calculation for a constant $w$ will yield 

\be
9w(w+1)+2(1+j)=0,
\label{weqe}
\ee

It is easy to see that $j=-1$ again gives two solutions of equation (\ref{weqe}) $w=0$ and $w=-1$ leading to a $\Lambda$CDM behaviour where $w=0$ corresponds to the cold dark matter and $w=-1$ corresponds to the cosmological constant. However, this will not lead to any interaction. One should also note that starting from the definition of the jerk parameter (equation (\ref{jerkparameter})), with the assumption that $j$ is a constant, one actually recovers one of the Einstein's equations for the system, where the nature of the matter sector depends on the value of the parameter. Thus the parameter $A$ is equivalent to the matter density parameter $\Omega_{m0}$ because for $j=-1$, the power of $(1+z)$ in the first term on the right hand side of equation (\ref{h2zAj}) is 3 and the second term is a constant, equivalent to the constant vacuum energy density. If the power of $(1+z)$ in the first term is different from  3, the dark matter is not separately conserved. This invokes the possibility of interaction between the dark matter and the dark energy. 

\par To investigate the nature of interaction for the present model, the total conservation equation which is a direct consequence of contracted Bianchi identity, can be divided into two parts as the followings,
\be 
\dot{\rho}_m+3H\rho_m=\eta,
\label{mattcon} 
\ee 
and 
\be 
\dot{\rho}_{DE}+3H(1+w_{DE})\rho_{DE}=-\eta.
\ee
The over head dots represent the differentiation with respect to cosmic time, $\rho_m$ and $\rho_{DE}$ are the matter density and dark energy density respectively, $w_{DE}$ is the dark energy equation of state parameter. As the dark energy and dark matter interact with themselves, they are not conserved individually. The growth rate of one component, namely $\eta$, is the decay rate of the other. In the present work, the possibility of interaction has been studied. For a dimensionless representation, $\eta$ has been scaled by ($3H_0^3/8\pi G$) and written as 
\be
Q=\frac{8\pi G}{3H_0^3}\eta.
\label{Qz}
\ee
For the $\Lambda$CDM model, the value of $Q$ is zero. In the present work, observational constraints on the late time evolution of $Q(z)$ has been obtained. As in the expression of the Hubble parameter (equation (\ref{h2zAj})), the first term of the right hand side, i.e. $A(1+z)^{\frac{3+\sqrt{9-8(1+j)}}{2}}$, is considered to be the matter density scaled by the present critical density, from equation (\ref{mattcon}), the interaction term $Q$ can be expressed in terms of the parameters and redshift, as
\be
Q(z)=A\Bigg(\frac{3-\sqrt{9-8(1+j)}}{2}\Bigg)(1+z)^{\frac{3+\sqrt{9-8(1+j)}}{2}}h(z).
\label{intQ}
\ee

It is important to note that the second  term on the right hand side of equation (\ref{h2zAj}) is considered to be the contribution from the dark energy density. Thus the expression of the dark energy equation of state parameter ($w_{DE}=p_{DE}/\rho_{DE}$) looks like,

\be
w_{DE}(z)=-\Bigg(\frac{3+\sqrt{9-8(1+j)}}{6}\Bigg)-\Big(\frac{A}{1-A}\Big)
\Bigg(\frac{3-\sqrt{9-8(1+j)}}{6}\Bigg)(1+z)^{\sqrt{9-8(1+j)}}.
\ee

\section{Results of statistical analysis}

\begin{figure}[htb]
\begin{center}
\includegraphics[angle=0, width=0.28\textwidth]{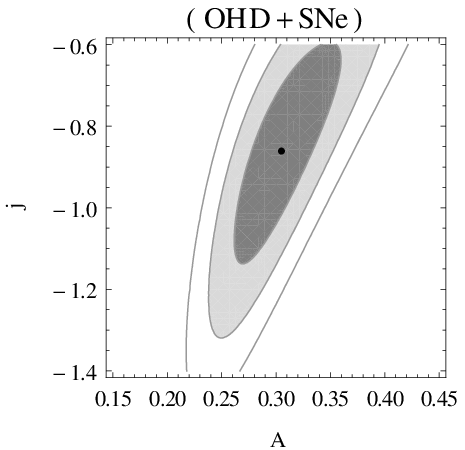}
\includegraphics[angle=0, width=0.28\textwidth]{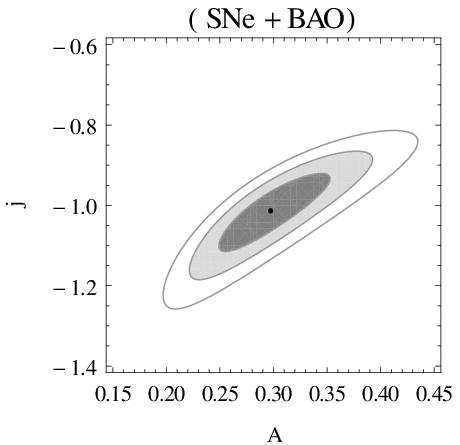}
\includegraphics[angle=0, width=0.28\textwidth]{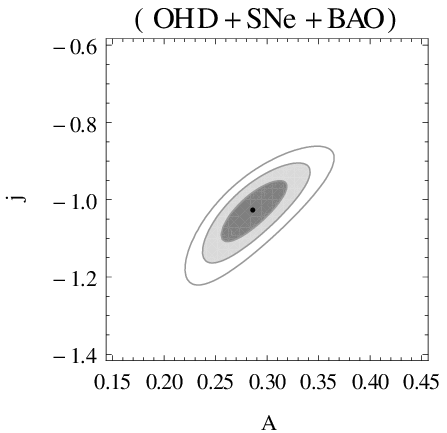}
\end{center}
\caption{{\small Confidence contours on the 2D parameter space of the reconstructed model. The 1$\sigma$, 2$\sigma$ and 3$\sigma$ confidence regions have been presented from inner to outer area and the central black dots represent the corresponding best fit point. The left panel shows the confidence contours obtained for the statistical analysis using OHD+SNe data, the middle panel is  obtained SNe+BAO and the right panel is for OHD+SNe+BAO.}}
\label{contourplot1}
\end{figure}

\begin{figure}[htb]
\begin{center}
\includegraphics[angle=0, width=0.35\textwidth]{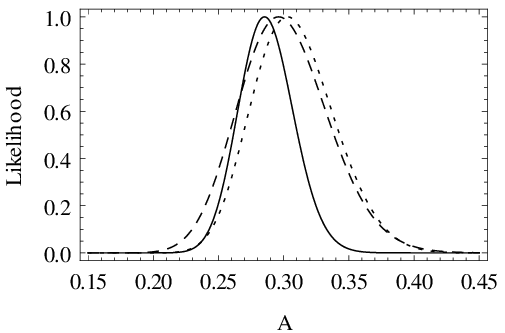}
\includegraphics[angle=0, width=0.35\textwidth]{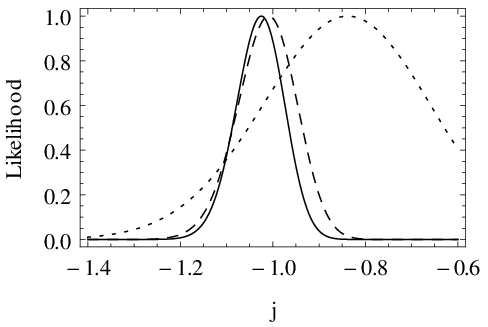}
\end{center}
\caption{{\small Plots of marginalized likelihood functions of the reconstructed model. The dotted curves represents the likelihood obtained for OHD+SNe, dashed curves represents the likelihood for SNe+BAO and the solid curves represents the likelihood for OHD+SNe+BAO.}}
\label{likelihoodplot1}
\end{figure}

Now the remaining task is to estimate the parameter values from observational data sets. In the present work, three different data sets have been adopted. These are (i) the observational Hubble parameter data (OHD), (ii) the distance modulus data of type Ia supernova (SNe) and (iii) baryon acoustic oscillation data (BAO). 

\par For the first set of data, measurement of Hubble parameter at different redshift by using differential age  of galaxies as an estimator \cite{simonohd}, measurement from red-enveloped galaxies \cite{sternohd}, measurement of expansion rate of the universe by Moresco et al \cite{morescoohd}, Measurement of expansion history from WiggleZ Dark Energy Survey as discussed by Blake et al. \cite{blakeOHD}, the measurement of Hubble parameter from Sloan Digital Sky Survey (SDSS) data by Zhang et al \cite{zhangohd} have been adopted in the present analysis. The recent measurement of Hubble parameterat z=2.34 by Delubac et al \cite{delubacohd} has also been incorporated in the present analysis. The present value of Hubble parameter $H_0$, estimated from the combined analysis with Planck+WP+highL+BAO \cite{planck}, has also been adopted in the present analysis.

\par  The supernova distance modulus data (SNe) of joint lightcurve analysis (jla) has been adopted in this work \cite{betoul} for the second category. The binned data of jla has been used along with the covariance matrix of the binning.

\par Finally, baryon acoustic oscillation (BAO) measurements at three different redshifts (6dF Galaxy Surve at redshift z=0.106 \cite{beutlerbao}, the Baryon Oscillation Spectroscopic Survey at redshift z=0.32 (BOSS LOWZ) and at redshift z=0.57 (BOSS CMASS) \cite{andersonbao}) along with the measurement of acoustic scale and comoving sound horizon at photon decoupling and drag epoch from CMB \cite{planck,wangwangCMB} have been adopted.

\par The $\chi^2$ minimization technique which is equivalent to the maximum likelihood analysis, has been adopted to find the best fit values of the model parameters. The results have been obtained for different combinations of the data sets. The $\chi^2$ is defined as 
\be 
\chi^2=\sum_i\frac{[\epsilon_{obs}(z_i)-\epsilon_{th}(z_i.\{\theta\})]^2}{\sigma_i},
\ee 
where $\epsilon_{obs}$ is the value of the observable measured at redshift $z_i$, $\epsilon_{th}$ from of the observable quantity as a function of the set of model parameters $\{\theta\}$ and $\sigma_i$ is the uncertainty associated to the measurement at $z_i$. The combined analysis has been carried out by adding the $\chi^2$ of the individual data sets taken into account for that particular combination. The combined $\chi^2$ is defined as,
\be
\chi^2_{combined}=\sum_d\chi^2_d,
\ee 
where $d$ represents the individual data set. 
\par The kinematic model discussed in the present work contains two parameters ($A$, $j$) where the parameter $A$ is coming as an integration constant and the $j$ is the constant jerk parameter. As mentioned earlier, it is imperative to note that the parameter $A$ is equivalent to matter density parameter. The value of jerk parameter $j$, estimated from observational data, would indicate the deviation, if any, of the present model from $\Lambda$CDM, for $j=-1$, the present model mimics the $\Lambda$CDM. 

\par The expression of the Hubble parameter  obtained for the present model (equation (\ref{h2zAj})) shows that the matter sector has two components. The first one, with constant coefficient $A$, is the dark matter density and the other one is the dark energy density. As the  energy density of relativistic particles, mainly the photon and neutrino, have an effective contribution to the dynamics of the of the universe at very high redshift, an additional energy density term, evolving as $(1+z)^4$ for radiation, has been taken into account while using the  angular diameter distance measurement in the analysis with BAO data. The present value of the energy density of relativistic particles scaled  by the present critical density is taken to be $\Omega_{r0}=9.2\times10^{-5}$ with the adopted fiducial value of current CMB temperature $T_0=2.7255$K. The adopted fiducial value of $T_0$ is based on the measurement of current CMB temperature $T_0=2.7255\pm0.0006$K \cite{fixsenT0}. As the prime endeavour of the present work is to reconstruct the late time dynamics of the universe, the radiation energy density is not taken in account as it has a negligible contribution at late time, i.e., in equation (\ref{h2zAj}).

\begin{table}[h!]
\begin{center}
\resizebox{0.6\textwidth}{!}{  
\begin{tabular}{ c |c |c c } 
\hline
 \hline
  Data & $\chi^2_{min}/d.o.f.$  & $A$ & $j$ \\ 
 \hline
  OHD+SNe & 47.30/54 & 0.305$\pm$0.023 & -0.861$\pm$0.127\\ 

  SNe+BAO & 33.95/28 & 0.297$\pm$0.024 & -1.014$\pm$0.045\\ 

  OHD+SNe+BAO & 48.28/54 & 0.286$\pm$0.015 & -1.027$\pm$0.037\\ 
 \hline
\hline
\end{tabular}
}
\end{center}
\caption{{\small  Results of statistical analysis with different combinations of the data sets. The value of $\chi^2_{min}/d.o.f.$ and the best fit values of the parameters along with the  associated 1$\sigma$ uncertainties are presented.}}
\label{tableAj}
\end{table}

\begin{figure}[htb]
\begin{center}
\includegraphics[angle=0, width=0.35\textwidth]{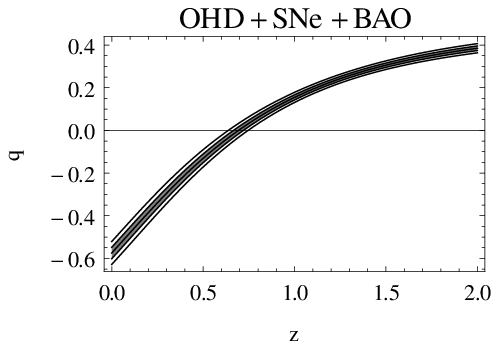}
\includegraphics[angle=0, width=0.35\textwidth]{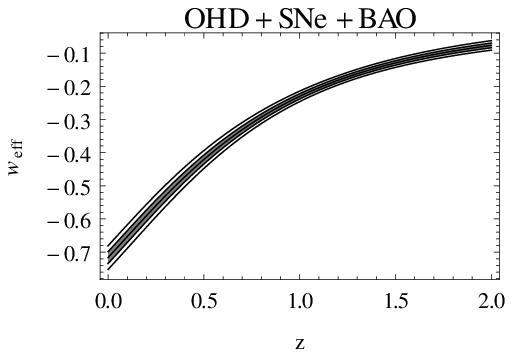}
\end{center}
\caption{{\small The plots of the deceleration parameter ($q(z)$) (left panel) and the effective equation of state parameter ($w_{eff}(z)$) (right panel) for the reconstructed model. The corresponding 1$\sigma$ and 2$\sigma$ confidence regions and the best fit curves obtained in the analysis combining OHD, SNe and BAO data sets, are presented.}}
\label{qzweffz}
\end{figure}

\begin{figure*}[ht]
\begin{center}
\includegraphics[angle=0, width=0.30\textwidth]{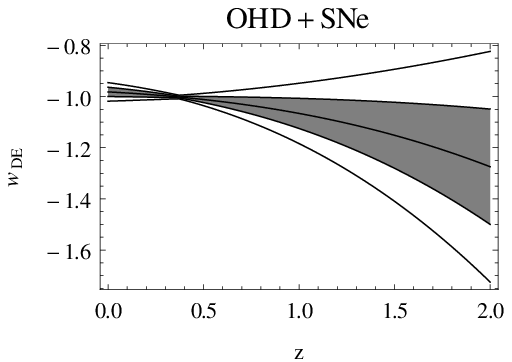}
\includegraphics[angle=0, width=0.30\textwidth]{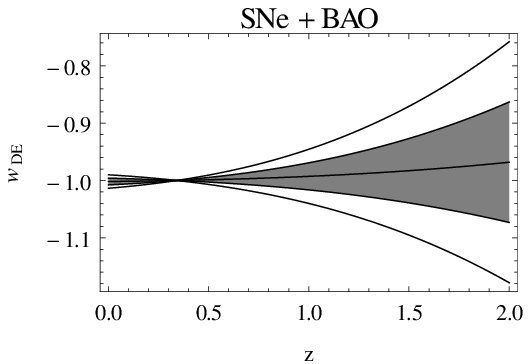}
\includegraphics[angle=0, width=0.30\textwidth]{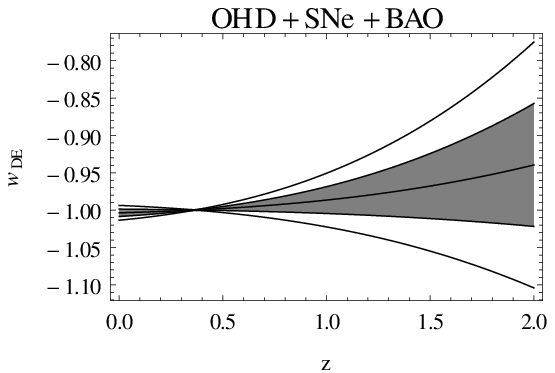}
\end{center}
\caption{{\small The plots of the dark energy equation of state parameter ($w_{DE}(z)$), obtained from the analysis with different combination of the data sets are presented. The corresponding 1$\sigma$ and 2$\sigma$ confidence regions and the best fit curves are shown.}}
\label{wDEz}
\end{figure*}

\begin{figure*}[htb]
\begin{center}
\includegraphics[angle=0, width=0.30\textwidth]{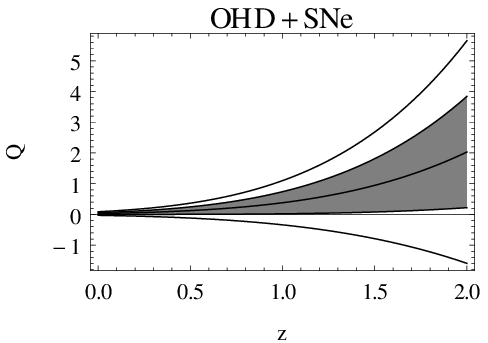}
\includegraphics[angle=0, width=0.30\textwidth]{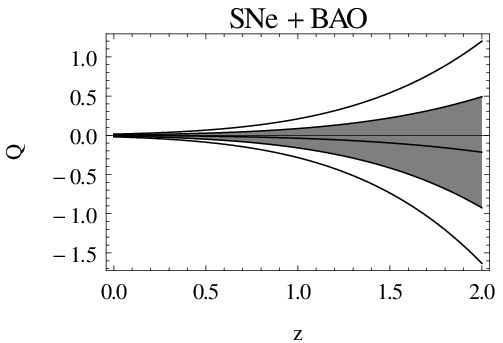}
\includegraphics[angle=0, width=0.30\textwidth]{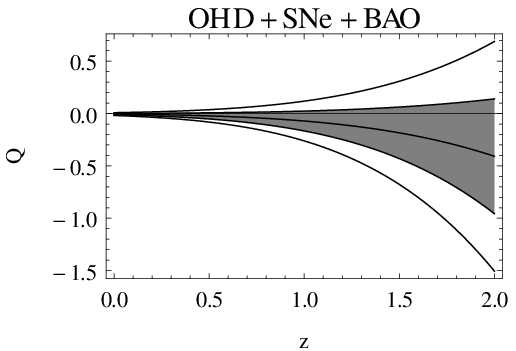}
\end{center}
\caption{{\small The plots of interaction term $Q(z)$, obtained from the analysis with different combination of the data sets are presented.  The corresponding 1$\sigma$ and 2$\sigma$ confidence regions and the best fit curves are shown. The Q=0 straight line represents the $\Lambda$CDM model.}}
\label{Qz}
\end{figure*}

\par Figure \ref{contourplot1} shows the confidence contours on the two dimensional (2D) parameter space of the model for different combinations of the data sets and figure \ref{likelihoodplot1} presents the plots of marginalized likelihoods as functions of the model parameters. The likelihoods are well fitted to Gaussian distributions. Table \ref{tableAj} shows the results of the statistical analysis for different combinations of the data sets. The reduced $\chi^2$ (i.e. $\chi^2/d.o.f.$) values are also presented to have an estimation of the goodness of fitting. Both figure \ref{contourplot1} and table \ref{tableAj} clearly show that the best fit value of $j$ is very close to $-1$, indicating clearly that the model with a constant jerk parameter is tantalizingly close to $\Lambda$CDM model.

\par Figure \ref{qzweffz} presents the plots of deceleration parameter $q(z)$ (left panel) and the effective or total equation of state $w_{eff}(z)$ (right panel) where $w_{eff}=p_{DE}/(\rho_m+\rho_{DE})$. The deceleration parameter plot clearly shows that the reconstructed model successfully generates the recent cosmic acceleration along with the decelerated expansion phase that prevailed in the past. The redshift of transition from decelerated to accelerated expansion phase obtained for the present model is 0.6 to 0.8 which is consistent with the result of recent analysis by Farooq and Ratra \cite{farooqratra}. Figure \ref{wDEz} presents the plots of dark energy equation of state parameter $w_{DE}$, obtained from the analysis with different combinations of the data sets. The best fit curves remain very close to the corresponding $\Lambda$CDM value $w_{DE}=-1$ and the deviation increases at high redshift. The best fit curve obtained in the analysis combining OHD, SNe and BAO data sets shows a slight inclination towards the non-phantom nature of dark energy. But the phantom nature ($w_{DE}<-1$) is also well within the 1$\sigma$ confidence region (figure \ref{wDEz}). The interaction between the dark matter and dark energy that crosses the phantom barrier of $w_{DE}=-1$ has been discussed by Guo {\it et al.} \cite{guopiao1,guopiao2}. It is clear from the plots the the nature of the best fit curve is sensitive to the combination of  the data sets taken into account.   The plots also show that the $w_{DE}(z)$ is constrained better at low redshift and the uncertainty increases at high redshift. It deserves mention in this context that the value of jerk parameter ($j$) and the the evolution of dark energy equation of state parameter ($w_{DE}$), obtained in the analysis with different combinations of the data sets,indicate that the reconstructed model is at close proximity of $\Lambda$CDM.

\begin{table}[h!]
\begin{center}
\resizebox{0.35\textwidth}{!}{  
\begin{tabular}{ c |c  } 
\hline
 \hline
  Data  & $Q(z=0)$ \\ 
 \hline
  OHD+SNe  & $0.0292\pm0.0293$\\ 

  SNe+BAO &  $-0.0026\pm0.0087$\\ 

  OHD+SNe+BAO &  $-0.0051\pm0.0067$\\ 
 \hline
\hline
\end{tabular}
}
\end{center}
\caption{{\small The present value of the interaction term i.e. $Q(z=0)$ obtained for different combinations of the data sets. The corresponding best fit values and the associated 1$\sigma$ uncertainties are presented.}}
\label{tableQ0}
\end{table}

\par Figure \ref{Qz} shows the evolution of  the interaction term $Q(z)$, defined in equation (\ref{intQ}). For the present model, any deviation from $\Lambda$CDM indicates a possibility of interaction between dark energy and dark matter. For non interacting models, the interaction term $Q(z)$ is zero. The plots of $Q(z)$ of the present model obtained from the analysis with different combination of the data sets show that the evolution of  $Q(z)$ is also sensitive to the choice of data sets. However, for all combinations of the data sets taken in the present work show the possibility of interaction between dark energy and dark matter is very low at present epoch. But the possibility of interaction is high at high redshift. The $\Lambda$CDM always remains within the 1$\sigma$ confidence region. The result obtained from the combination of SNe, OHD  shows a higher preference towards the interaction at high redshift (left panel of figure \ref{Qz}). But the addition of BAO data brings best fit curve closer to $\Lambda$CDM. It is also easy to note that the present analysis allows both positive and negative value for the interaction term $Q(z)$. Table \ref{tableQ0} presents the present values of the interaction term obtained from the analysis with different combination of the data sets. The result obtained from the analysis combining SNe, OHD and BAO data shows that the best fit curve of $Q(z)$ has an inclination towards negative value. But the possibility of a positive $Q(z)$ is also well within the 1$\sigma$ confidence region. The requirement of a positive $Q(z)$ in the context of thermodynamics has been discussed by Pavon and Wang \cite{pavonwang}.

\section{Discussion}
The present work is an attempt to search for the possibility of interaction between the dark matter and the so-called dark energy with a kinematic approach. The other crucial factor is that we start from the dimensionless jerk parameter $j$ which is a third order derivative of the scale factor $a$. This choice is of a natural interest, as the evolution of $q$, the second order time derivative of $a$ is an observational quantity now. We start from the geometrical definition of jerk, and do not use even Einstein equations to start with. We reiterate the word of caution, the conclusion that any deviation of the jerk parameter from $-1$ indicates an interaction between dark matter and dark energy sectors is actually based on the particular choice of identification of the matter density term in equation (\ref{h2zAj}).  

\par The result obtained clearly shows that the best fit value of $j$ (chosen as a constant parameter) is very close to $-1$, which is consistent with a $\Lambda$CDM model. The interaction term $Q$, in a dimensionless representation, is very close to zero at the present era. This is completely consistent as $\Lambda$, being a constant, does not exchange energy with dark matter. Table \ref{tableQ0} shows the best fit values of $Q$ at $z=0$ for various combinations of data sets. It is easily seen that the values are two orders of magnitude smaller than $\Omega_{m0}$ and $\Omega_{DE0}$, which are approximately $0.3$ and $0.7$ respectively. All these quantities are expressed in a dimensionless way. So this comparison is possible.

\par As already mentioned, investigations regarding a reconstruction of interaction are not too many. But the very recent work by Yang, Guo and Cai\cite{yangguocai} is a rigorous and elaborate one. The method adopted is the Gaussian processes. Although the work is model dependent, as the equation of state parameter is not specified, it can be applied to a large variety of dark energy models. The $w$CDM model is particularly emphasized. The basic result is the same as that of the present work, the interaction appears to be negligible and consistent with the $\Lambda$CDM model at $z=0$.

\par An intriguing feature in both the present work and that in ref \cite{yangguocai}, is that although the best fit value still hovers around being negligible, it is allowed to have a non-trivial value for higher $z$ even in the 1$\sigma$ confidence region. So the interaction, if any, took place in the earlier epoch. The physics of this is not yet quite understood.

\par Another interesting result in the present work is the fact that $Q$, if it has a sizable value, it can be both positive or negative, so the pumping of energy is possible both ways. Intuitively it might appear that the dark energy should grow at the expense of dark matter ($Q<0$). However, the thermodynamic considerations demand that the flow of energy should be the other way round, from dark energy to dark matter\cite{pavonwang}.

\par It  has already been mentioned that the reconstructed model mimics the $\Lambda$CDM for the value of cosmological jerk parameter $j=-1$ and prevent the possibility of interaction between dark energy and dark matter. Any observational measurement which is based on the fiducial assumption of a $\Lambda$CDM cosmology, might affects the results of statistical analysis by making the parameter values highly biased towards the corresponding $\Lambda$CDM values and leading to far too optimistic error bars. Hence such kind of data, like the CMB distance prior measurement, has not been introduced directly in the likelihood analysis. The correlations of distance modulus measurement of type Ia supernova have been taken into account as it might  have its signature on the results. The error bars of dark energy equation of state parameter ($w_{DE}$) are found to be quite consistent with the results of Planck 2015 \cite{planck2015}. 
  
\par The recent works on the reconstruction of jerk show that definitely a $\Lambda$CDM is favoured\cite{zhai, mukban}, where $j$ was allowed to be a function of $z$. Now a reconstruction of interaction through a constant $j$ also indicates towards a similar result.

\end{document}